\documentclass{article}
\usepackage{spconf}
\usepackage{amsmath}
\usepackage{graphicx}  
\usepackage{amssymb}
\usepackage{multirow}
\usepackage{threeparttable}
\usepackage{cite}
\usepackage{booktabs}
\usepackage{makecell}
\usepackage{diagbox}
\usepackage{color}

\def\Vec#1{\textit{\boldmath $#1$}}

\def\red#1{#1}

\title{Multi-ACCDOA: Localizing and Detecting Overlapping Sounds from the Same Class with Auxiliary Duplicating Permutation Invariant Training}
\makeatletter
\def\@name{ \emph{Kazuki Shimada, Yuichiro Koyama, Shusuke Takahashi, Naoya Takahashi, Emiru Tsunoo, Yuki Mitsufuji}}
\makeatother

\address{Sony Group Corporation, Tokyo, Japan}
\begin{document}
\ninept

\maketitle

\begin{abstract}

Sound event localization and detection~(SELD) involves identifying the direction-of-arrival~(DOA) and the event class.
The SELD methods with a class-wise output format make the model predict activities of all sound event classes and corresponding locations.
The class-wise methods can output activity-coupled Cartesian DOA~(ACCDOA) vectors, which enable us to solve a SELD task with a single target using a single network.
However, there is still a challenge in detecting the same event class from multiple locations.
To overcome this problem while maintaining the advantages of the class-wise format, we extended ACCDOA to a multi one and proposed auxiliary duplicating permutation invariant training~(ADPIT).
The multi-ACCDOA format (a class- and track-wise output format) enables the model to solve the cases with overlaps from the same class.
The class-wise ADPIT scheme enables each track of the multi-ACCDOA format to learn with the same target as the single-ACCDOA format.
In evaluations with the DCASE 2021 Task 3 dataset, the model trained with the multi-ACCDOA format and with the class-wise ADPIT detects overlapping events from the same class while maintaining its performance in the other cases.
Also, the proposed method performed comparably to state-of-the-art SELD methods with fewer parameters.

\end{abstract}


\section{Introduction}
\label{sec:intro}

Sound event localization and detection~(SELD) involves identifying the direction-of-arrival~(DOA) and the type of sound events.
SELD has played an essential role in many applications, such as surveillance~\cite{crocco2016audio,valenzise2007scream}, bio-diversity monitoring~\cite{chu2009environmental}, and context-aware devices~\cite{yalta2017sound,sun2021emergency}.
Recent competitions such as the DCASE challenge show significant progress in the SELD research area using neural-network~(NN)-based methods~\cite{politis2020overview}.

NN-based SELD methods can be categorized into two output formats.
The first is the class-wise output format, in which the model predicts activities of all event classes and corresponding locations~\cite{adavanne2018sound,politis2020dataset,wang2021four,cao2019polyphonic,shimada2021accdoa,politis2021dataset,emmanuel2021multi}.
Adavanne {\it et al.} proposed SELDnet, which detects sound events and estimates the corresponding DOAs using two branches: an sound event detection~(SED) branch and a DOA branch~\cite{adavanne2018sound,politis2020dataset}.
Activity-coupled Cartesian DOA~(ACCDOA) assigns an event activity to the length of a corresponding Cartesian DOA vector~\cite{shimada2021accdoa,politis2021dataset}, which enables a SELD task to be solved without branching.
In the evaluations on the SELD task for DCASE 2020 Task 3, the single-ACCDOA format outperformed the two-branch format with fewer parameters~\cite{shimada2021accdoa}.
The second is the track-wise output format, where each track detects one event class and a corresponding location~\cite{cao2021improved,nguyen2020sequence,he2021sounddet}.
Cao {\it et al.} proposed an event independent network V2 (EINV2), which decomposes the SELD output into event-independent tracks~\cite{cao2021improved}.
They incorporate permutation invariant training~(PIT)~\cite{kolbaek2017multitalker} into a SELD task to solve a track permutation problem, in which an event cannot be fixedly assigned to a track.

While the track-wise format enables the model to detect the same event class from multiple locations, the track-wise format only detects one event class and a corresponding location in each track~\cite{cao2021improved}.
ACCDOA vectors can be used only in the class-wise methods.
Therefore, the track-wise format cannot exploit the advantage of the ACCDOA format, which enable us to solve a SELD task with a single target using a single network~\cite{shimada2021accdoa}.
On the other hand, there is still a challenge in detecting overlapping events from the same class because the class-wise format assigns only one location to each event class.

\begin{figure}[t]
    \centering
    \centerline{\includegraphics[width=0.80\linewidth]{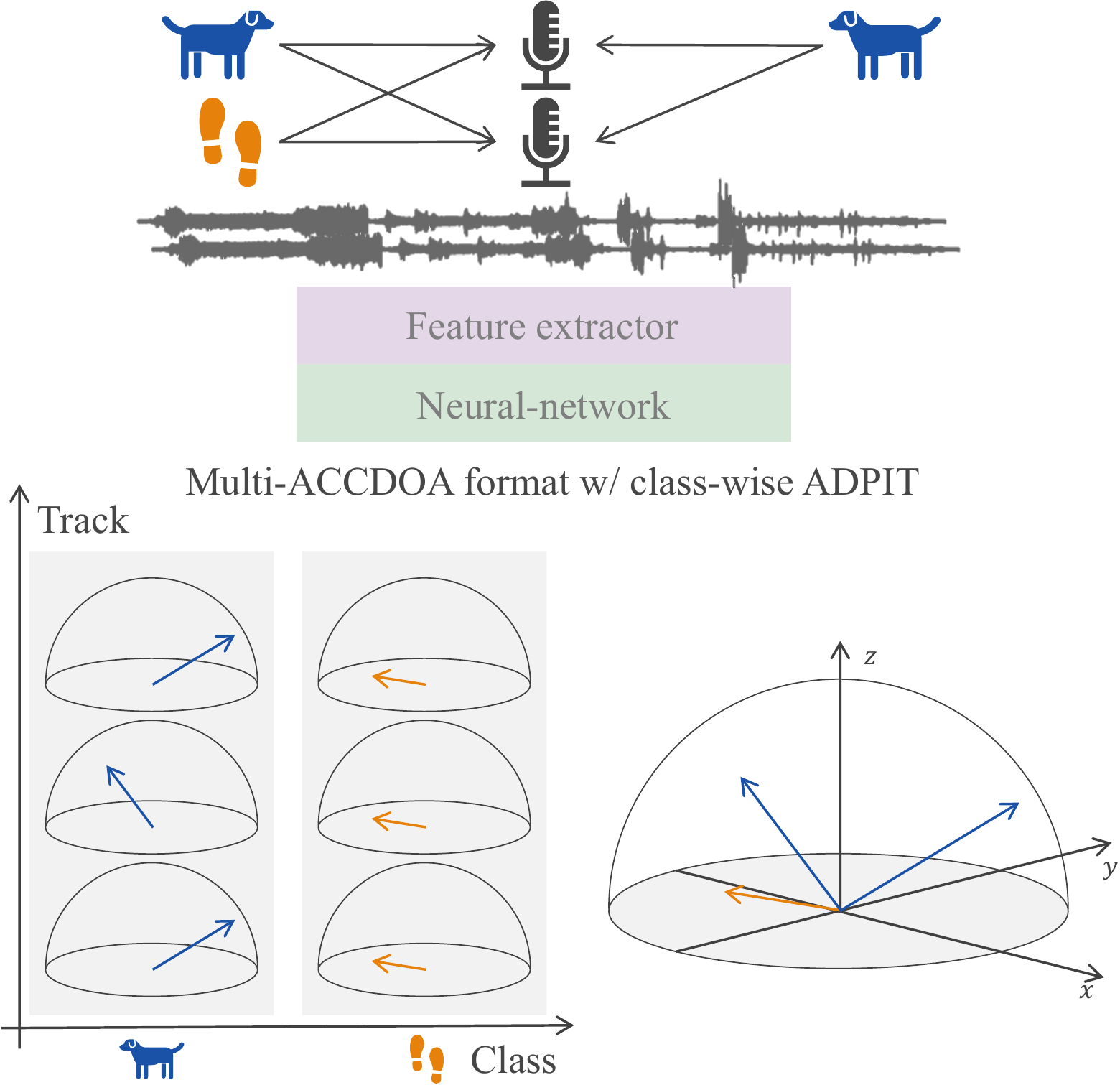}}
    \vspace{-2mm}
    \caption{
    The multi-ACCDOA format enables the model to detect and localize overlapping sounds from the same class, e.g., overlapping barking dogs.
    We show a 3-track 2-class multi-ACCDOA format in the bottom left.
    The ACCDOA vectors are trained to indicate the DOA of the sounds using the class-wise ADPIT.
    During inference, we unify duplicated outputs into the final output of the multi-ACCDOA format in the bottom right.
    }
    \label{fig:overview}
    \vspace{-5mm}
\end{figure}

To overcome the overlap problem while maintaining the advantages of the class-wise format, we extended the single-ACCDOA format to a multi-ACCDOA format (a class- and track-wise output format).
We also proposed auxiliary duplicating permutation invariant training~(ADPIT), inspired by auxiliary autoencoding permutation invariant training~(A2PIT) on a source separation task with a variable number of sources~\cite{luo2020separating}.
The class-wise ADPIT scheme enables us to solve the track permutation problem with duplicated ACCDOA vectors, which can help the model maintain the performance in the cases with no overlaps from the same class.
Fig.~\ref{fig:overview} shows the overview of the proposed method.
In experimental evaluations using the DCASE 2021 Task 3 dataset~\cite{politis2021dataset}, the result shows the multi-ACCDOA method with the class-wise ADPIT detected overlapping events from the same class without any degradation in the other cases.
Also, the proposed method performed comparably to state-of-the-art methods with fewer parameters.

\section{Related work}
\label{sec:related}


The class-wise methods can output ACCDOA vectors, which assign an activity to the length of a corresponding Cartesian DOA vector~\cite{shimada2021accdoa}.
When a class is active, an ACCDOA vector is set to be a unit vector; otherwise, it is set to be a zero vector.
That enables the model to solve a SELD task with a single objective such as the mean squared error~(MSE) using a single network.
As the single-ACCDOA format outperforms the two-branch format in the joint SELD metrics with a simpler architecture~\cite{shimada2021accdoa}, ACCDOA was adopted to the DCASE 2021 Task 3 baseline~\cite{politis2021dataset}.
A convolutional recurrent neural-network~(CRNN), consisting of convolution blocks, gated recurrent unit~(GRU) layers, and a fully-connected layer, is used to estimate ACCDOA vectors in~\cite{shimada2021accdoa}.

EINV2 uses the track-wise output format, where each event is exclusively assigned to one track~\cite{cao2021improved}.
That induces the track permutation problem, in which an event cannot be fixedly assigned to a track.
To solve the problem, PIT~\cite{kolbaek2017multitalker} is incorporated into a SELD task~\cite{cao2021improved}.
PIT assigns targets to different tracks to constitute all possible permutations.
Then the loss for each permutation is calculated.
The lowest loss is selected as the actual loss.
In the evaluation in~\cite{cao2021improved}, the frame-level PIT outperformed the chunk-level PIT.
The network architecture consists of convolution blocks, multi-head self-attention~(MHSA) blocks, fully connected layers, and soft parameter sharing~(PS) between an SED branch and a DOA branch~\cite{cao2021improved}.

A2PIT is used in the source separation task with a variable number of sources~\cite{luo2020separating}.
With the scheme, the mixture signal itself is used as the auxiliary targets instead of low-energy random noise signals when there are fewer targets than tracks~\cite{luo2020separating}.
That enables the model to use any objective functions such as scale-invariant signal-to-distortion ratio~(SI-SDR).
The idea of replacing auxiliary targets inspires us to use another auxiliary target instead of a zero vector in the SELD task.
A similarity threshold between the mixture and the outputs was used during inference to determine valid outputs~\cite{luo2020separating}.

\section{Proposed Method}
\label{sec:p_method}

We formulate the multi-ACCDOA format and show how to train the model with the format using the class-wise ADPIT.
Then we describe the inference step of the proposed method.

\subsection{Multi-ACCDOA format}
\label{ssec:m_accdoa}

The multi-ACCDOA format is an extension of the single-ACCDOA format~\cite{shimada2021accdoa} to a track dimension; thus, it is a class- and track-wise format.
While each track in the track-wise format only detects one event class and a corresponding location, each track in the multi-ACCDOA format is equivalent to the single-ACCDOA format, where a track predicts activities of all target classes and their corresponding locations.

The $N$-track $C$-class $T$-frame multi-ACCDOA format, $\Vec{P} \in {\mathbb{R}}^{3 \times N \times C \times T}$, is formulated with activities and Cartesian DOAs.
Each sound event class of a track is represented by three nodes corresponding to the sound event location in the $x$, $y$, and $z$ axes.
Let $\Vec{a} \in {\mathbb{R}}^{N \times C \times T}$ be activities, whose reference value is ${a}_{nct}^{*} \in \{0, 1\}$, i.e., it is 1 when the event is active and 0 when inactive.
$n, c, t$ indicates an output track number, a target class, and a time frame.
Also, let $\Vec{R} \in {\mathbb{R}}^{3 \times N \times C \times T}$ be Cartesian DOAs, where the length of each Cartesian DOA is 1, i.e., $||\Vec{R}_{nct}|| = 1$ when a class $c$ is active.
$||\cdot||$~is the L2 norm.
An ACCDOA vector in the multi-ACCDOA format is formulated as follows:
\begin{align}
    \Vec{P}_{nct} =
        {a}_{nct} \Vec{R}_{nct}.
    \label{eq:m_accdoa}
\end{align}
In this study, to estimate the ACCDOA vectors in the multi-ACCDOA format $\Vec{P}_{nct}$, the CRNN architecture is used as in~\cite{shimada2021accdoa}.

\subsection{Class-wise auxiliary duplicated PIT}
\label{ssec:adpit}

\begin{figure}[t]
    \centering
    \centerline{\includegraphics[width=0.99\linewidth]{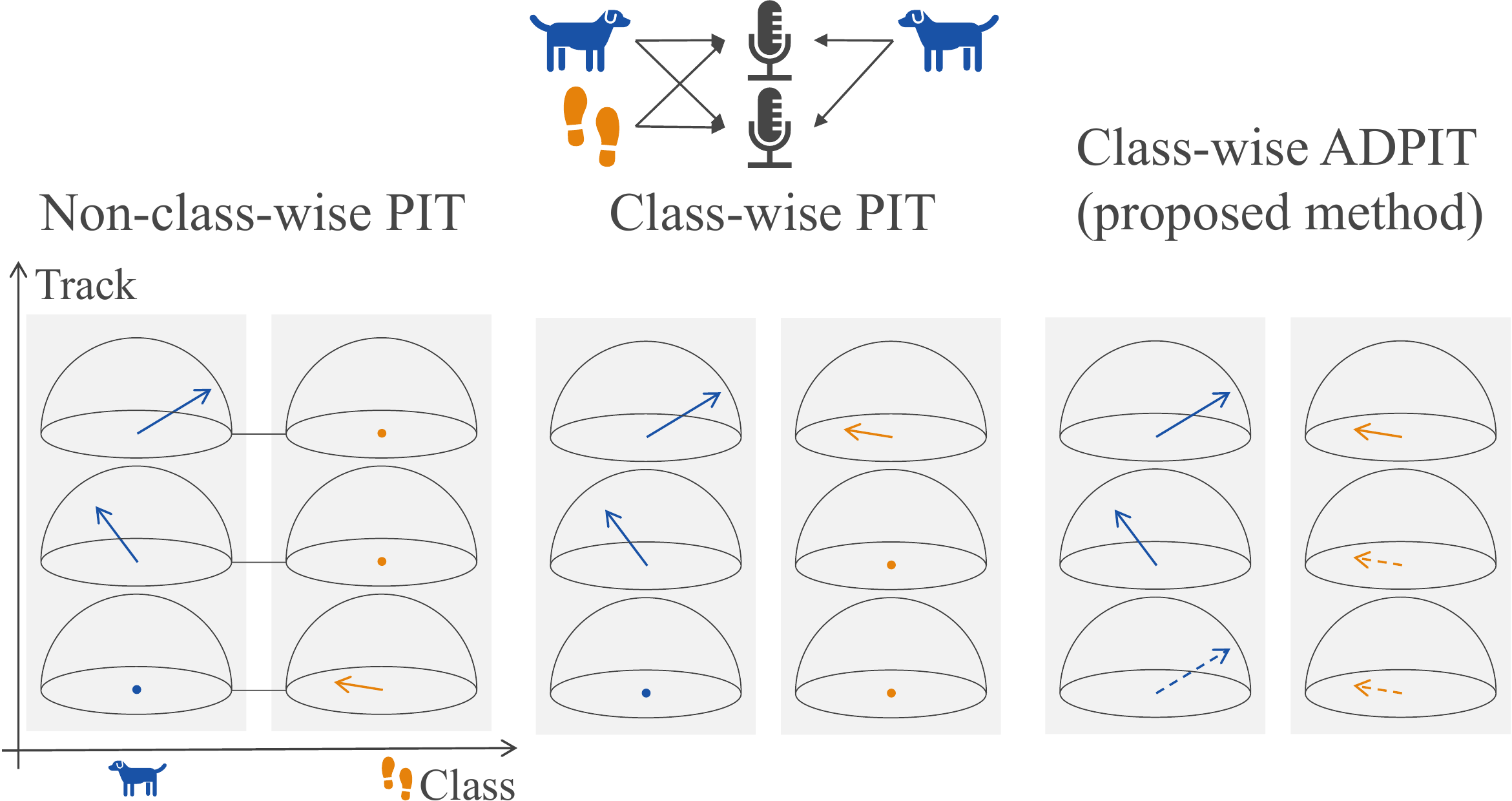}}
    \vspace{-2mm}
    \caption{
    We show a possible permutation for each PIT with a case illustrated at the top.
    There are two barking dogs and one footstep in a time frame.
    We explain a 3-track 2-class multi-ACCDOA format with three frame-level PITs: a non-class-wise PIT, a class-wise PIT, and a class-wise ADPIT.
    A solid line shows an original target, and a dot shows a zero vector.
    A dotted line depicts a duplicated target.
    The non-class-wise PIT exclusively assigns one event to one track regardless of the event class.
    The class-wise PIT assigns one event to one track for each class.
    Zero vectors are assigned to the other tracks.
    With the class-wise ADPIT, each track is trained to output an original target or a duplicated target.
    }
    \label{fig:adpit_a_possible}
    \vspace{-5mm}
\end{figure}

Similarly to the track-wise approaches, the multi-ACCDOA format also suffers from the track permutation problem.
To overcome this issue, we adopt PIT for the training process.
The frame-level PIT~\cite{cao2021improved} is used in this study.
Assume all possible permutations constitute a permutation set $\mathrm{Perm}$.
$\alpha \in \mathrm{Perm}(t)$ is one possible frame-level permutation at frame $t$.
A PIT loss for the multi-ACCDOA format can be straightforwardly written as follows:
\begin{align}
    \mathcal{L}^\mathrm{PIT} &= \frac{1}{T} \sum_{t}^{T} \min_{\alpha \in \mathrm{Perm}(t)} {l}^{\mathrm{ACCDOA}}_{\alpha, t}, 
    \label{eq:loss_frame_adpit} \\
    {l}^\mathrm{ACCDOA}_{\alpha, t} &= \frac{1}{NC} \sum_{n}^{N} \sum_{c}^{C} \mathrm{MSE}(\Vec{P}_{\alpha, nct}^{*}, \hat{\Vec{P}}_{nct}),
    \label{eq:loss_frame_accdoa}
\end{align}
where $\Vec{P}_{\alpha, nct}^{*}$ is an ACCDOA target of a permutation $\alpha$, and $\hat{\Vec{P}}_{nct}$ is an ACCDOA prediction at track $n$, class $c$, and frame $t$.
We refer this PIT as a non-class-wise PIT because the PIT loss in Eq.~(\ref{eq:loss_frame_accdoa}) exclusively assigns one sound event to one track regardless of the sound event class~\cite{cao2021improved}.
We use MSE as a loss function for the multi-ACCDOA format~\cite{shimada2021accdoa}.
The left in Fig.~\ref{fig:adpit_a_possible} shows a possible permutation of a 3-track 2-class multi-ACCDOA format with the non-class-wise PIT.
The blue and orange vectors indicate ACCDOA vectors of a barking dog and a footstep, respectively.


Because the multi-ACCDOA format has a class dimension, it is reasonable to set a permutation for each class.
Therefore, we extend the non-class-wise PIT to a class-wise formulation, namely class-wise PIT.
$\beta \in \mathrm{Perm}(ct)$ is one possible class-wise frame-level permutation at class $c$ and frame $t$.
The class-wise PIT loss for the multi-ACCDOA format can be written as follows:
\begin{align}
    \mathcal{L}^\mathrm{PIT} &= \frac{1}{CT} \sum_{c}^{C} \sum_{t}^{T} \min_{\alpha \in \mathrm{Perm}(ct)} {l}^{\mathrm{ACCDOA}}_{\alpha, ct}, 
    \label{eq:loss_adpit} \\
    {l}^\mathrm{ACCDOA}_{\alpha, ct} &= \frac{1}{N} \sum_{n}^{N} \mathrm{MSE}(\Vec{P}_{\alpha, nct}^{*}, \hat{\Vec{P}}_{nct}).
    \label{eq:loss_accdoa}
\end{align}
The center in Fig.~\ref{fig:adpit_a_possible} shows a permutation example for the class-wise PIT, where a permutation is set for each class.

The class-wise PIT assigns one active event to only one track for each class.
When there are fewer active events than tracks in a event class, we need to assign zero vectors as auxiliary targets for the inactive tracks in the class.
The class-wise PIT cannot enable every track of the multi-ACCDOA format to learn with the same target as the single one, i.e., a unit vector.
As a result, the auxiliary targets for the class-wise PIT, i.e., zero vectors, might interfere the optimization of multi-ACCDOA training.

To enable every track of the multi-ACCDOA format to learn with the same target as the single one, we duplicate the original target as the auxiliary targets of the other tracks instead of zero vectors.
We refer to the PIT framework with the duplicated targets as ADPIT.
The multi-ACCDOA format tries to output ${M}_{ct} (\leq N)$ original targets and $N - {M}_{ct}$ duplicated targets in class ${c}$ and frame ${t}$, and \red{the PIT framework} is applied to find the best permutations.
The class-wise ADPIT loss is the same formula as the class-wise PIT in Eq.~(\ref{eq:loss_adpit}).
The difference lies in their permutation set.
We show a possible permutation in the right side in Fig.~\ref{fig:adpit_a_possible} for the class-wise ADPIT.
Each track is trained with an original target or a duplicated target.

\begin{figure}[t]
    \centering
    \centerline{\includegraphics[width=0.95\linewidth]{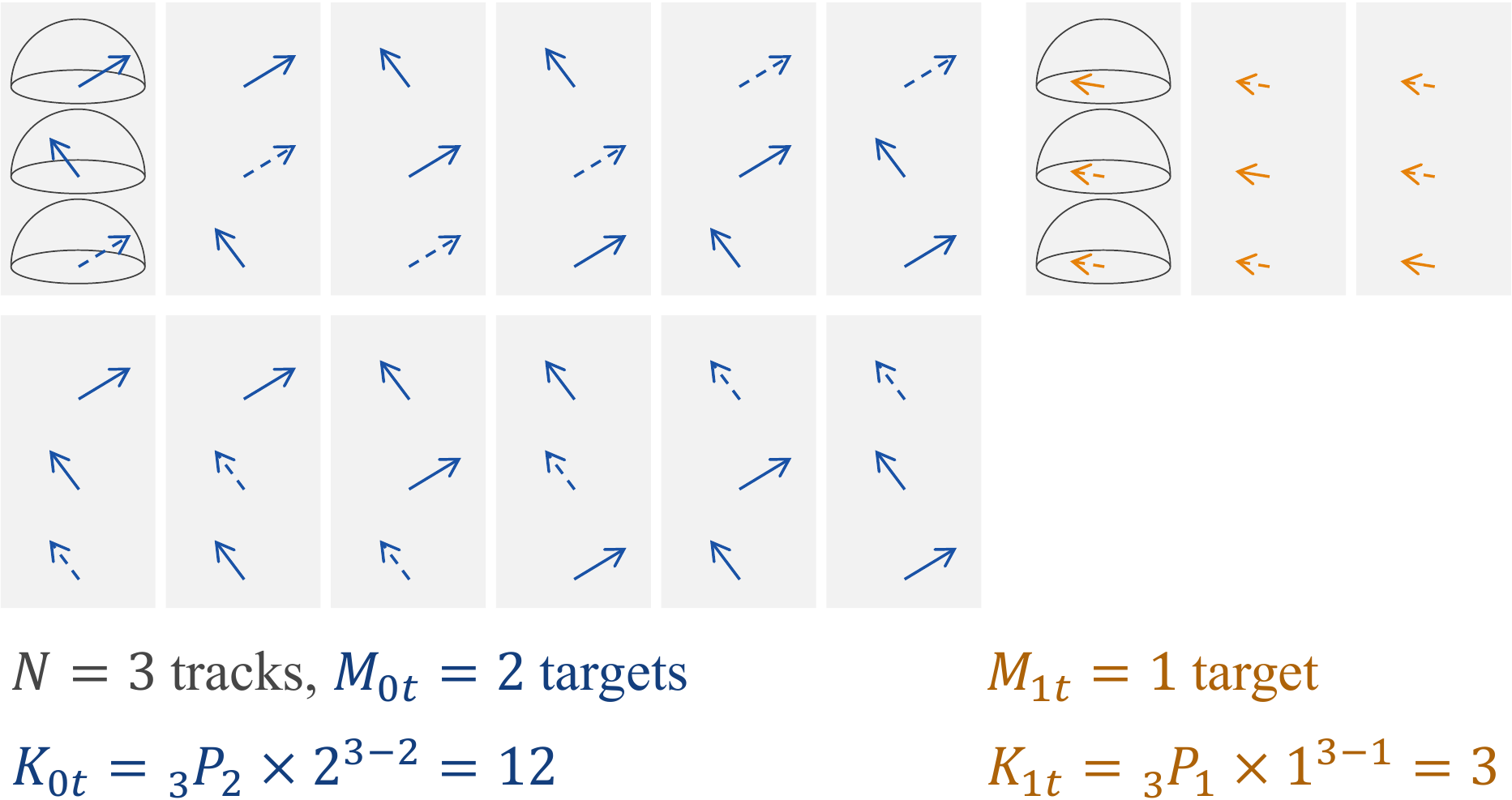}}
    \vspace{-3mm}
    \caption{
    We show all permutations for the class-wise ADPIT in the multi-ACCDOA format with the case in the top of the Fig.~\ref{fig:adpit_a_possible}.
    The number of tracks and classes are three and two, i.e., $N = 3, C = 2$.
    There are two barking dogs and one footstep in a time frame $t$, i.e., ${M}_{0t} = 2, {M}_{1t} = 1$.
    Each permutation consists of all original targets and some duplicated targets.
    The possible number of permutations, ${K}_{ct}$, is calculated with Eq.~(\ref{eq:num_pernumtation}).
    }
    \label{fig:adpit_all}
    \vspace{-5mm}
\end{figure}

The possible number of permutations for the class-wise ADPIT in the permutation set, ${K}_{ct}$, is calculated as follows:
\begin{align}
    {K}_{ct} = 
    \begin{cases}
        {}_{N} \mathrm{P}_{{M}_{ct}} \times {{M}_{ct}}^{N - {M}_{ct}} & \text{\red{if ${M}_{ct} > 0$,}} \\
        \red{1}  & \text{\red{if ${M}_{ct} = 0$,}}
    \end{cases}
    \label{eq:num_pernumtation}
\end{align}
\red{where ${}_{N} \mathrm{P}_{{M}_{ct}}$ denotes ${M}_{ct}$-permutations of $N$.}
Fig.~\ref{fig:adpit_all} shows all permutations for the class-wise ADPIT.
In practice, original targets and duplicated targets cannot be distinguished.
Therefore, the number of actual permutations is smaller than ${K}_{ct}$.

\subsection{Unification of duplicated outputs during inference}
\label{ssec:unify}

After the training with the class-wise ADPIT, during inference, duplicated outputs are unified in three steps.
First, we use threshold processing to determine whether each output is active or not.
Second, if there are two or more active outputs from the same class, we calculate the similarity between the active outputs.
In this study, we use angle difference as the similarity.
Third, if an angle difference between the outputs is smaller than a threshold, i.e., the outputs are similar, we take the average of the outputs to a unified output.

\section{Experimental evaluation}
\label{sec:exp}

\begin{table*}[t]
    \centering
    \caption{SELD performance in the single- and multi-ACCDOA formats evaluated for the testing fold of the DCASE 2021 Task 3 development set. All formats used the same CRNN architecture. Th means threshold, and ov means overlap. $\rm{{F}_{20^{\circ}}}$ and $\rm{{LR}_{CD}}$ are reported in percentages.}
    \scalebox{0.86}{
        \begin{tabular}{lc|c|ccccc|cc}
        \toprule
        & Th of & \# of & & & & & & $\rm{{LR}_{CD}}$ w/ ov & $\rm{{LR}_{CD}}$ \\
        Format & unification & params & $\rm{{ER}_{20^{\circ}}}$ & $\rm{{F}_{20^{\circ}}}$ & $\rm{{LE}_{CD}}$ & $\rm{{LR}_{CD}}$ & $\rm{\mathcal{E}_{SELD}}$ & from the same class & w/o \\
        \midrule
        Single-ACCDOA & - & 5.89 M & 0.588 & 54.2 & $18.8^{\circ}$ & 62.0 & 0.382 & 50.7 & 64.5 \\
        Multi-ACCDOA w/ non-class-wise PIT & - & 5.93 M & 0.650 & 41.6 & $23.5^{\circ}$ & 52.7 & 0.459 & 51.6 & 52.9 \\
        Multi-ACCDOA w/ class-wise PIT & - & 5.93 M & 0.671 & 43.7 & $23.7^{\circ}$ & 59.2 & 0.443 & 57.1 & 59.7 \\
        Multi-ACCDOA w/ class-wise ADPIT (proposed) & $15^{\circ}$ & 5.93 M & 0.596 & 55.3 & {\bf 18.4}$^{\circ}$ & {\bf 64.4} & 0.375 & {\bf 57.9} & {\bf 65.6} \\
        & $30^{\circ}$ & 5.93 M & 0.586 & {\bf 55.5} & {\bf 18.4}$^{\circ}$ & 63.9 & {\bf 0.374} & 55.5 & {\bf 65.6} \\
        & $45^{\circ}$ & 5.93 M & {\bf 0.584} & 55.4 & $18.5^{\circ}$ & 63.7 & {\bf 0.374} & 54.6 & {\bf 65.6} \\
        \bottomrule
        \end{tabular}
    }
    \label{tb:result_accdoa}
    \vspace{-4mm}
\end{table*}

\begin{table*}[t]
    \centering
    \caption{SELD performance in the multi-ACCDOA format and the track-wise format evaluated for the testing fold of the DCASE 2021 Task 3 development set. While the multi-ACCDOA format used the CNN-MHSA, the track-wise format used the original EINV2 architecture.}
    \scalebox{0.86}{
        \begin{tabular}{lc|c|ccccc|cc}
        \toprule
        & Th of & \# of & & & & & & $\rm{{LR}_{CD}}$ w/ ov & $\rm{{LR}_{CD}}$ \\
        Format & unification & params & $\rm{{ER}_{20^{\circ}}}$ & $\rm{{F}_{20^{\circ}}}$ & $\rm{{LE}_{CD}}$ & $\rm{{LR}_{CD}}$ & $\rm{\mathcal{E}_{SELD}}$ & from the same class & w/o \\
        \midrule
        Multi-ACCDOA w/ class-wise ADPIT (proposed) & $30^{\circ}$ & 6.85 M & {\bf 0.635} & {\bf 55.3} & 18.5$^{\circ}$ & 67.0 & {\bf 0.379} & 56.9 & 69.2 \\
        Track-wise w/ PIT & - & 22.0 M & 0.676 & 55.0 & {\bf 17.4}$^{\circ}$ & {\bf 69.1} & 0.383 & {\bf 62.6} & {\bf 70.4} \\
        \bottomrule
        \end{tabular}
    }
    \label{tb:result_einv2}
    \vspace{-4mm}
\end{table*}

We evaluated the multi-ACCDOA format trained with three PITs in Sec.~\ref{sec:p_method}, the single-ACCDOA format, and the track-wise format using TAU Spatial Sound Events 2021~\cite{politis2021dataset}.
The proposed method is also compared with state-of-the-art SELD methods.

\subsection{Task setups}
\label{ssec:tasks}

We used the development set of TAU Spatial Sound Events 2021 - Ambisonic with the suggested setup for DCASE 2021 Task 3~\cite{politis2021dataset}.
The dataset contained 600 one-minute sound scene recordings: 400 for training, 100 for validation, and 100 for testing.
The sound scene recordings were synthesized by adding sound event samples convolved with room impulse response~(RIR) to spatial ambient noise.
The sound event samples consisted of 12 event classes such as footsteps and a barking dog.
The RIRs and ambient noise recordings were collected at 15 different indoor locations.
Each event had an equal probability of being either static or moving.
The moving sound events were synthesized with 10, 20, or 40 degrees per second.
Up to three overlapping sound events are possible, temporally and spatially.
In addition, there are simultaneous directional interfering sound events with their temporal activities, static or moving.
Signal-to-noise ratios ranged from 6 to 30 dB.

Following the setup, five metrics were used for the evaluation~\cite{mesaros2019joint}.
The first was the localization error ${LE}_{CD}$, which expresses the average angular distance between the same class's predictions and references.
The second was a simple localization recall metric ${LR}_{CD}$, which tells the true positive rate of how many of these localization estimates were detected in a class out of the total number of class instances.
The following two metrics were the location-dependent error rate ${ER}_{20^{\circ}}$ and F-score ${F}_{20^{\circ}}$,
 where predictions were considered true positives only when the distance from the reference was less than  $20^{\circ}$.
We also adopted an aggregated SELD error, $\rm{\mathcal{E}_{SELD}}$, which is defined as
\begin{align}
    \rm{\mathcal{E}_{SELD}} = \frac{\rm{{ER}_{20^{\circ}}} + ( 1 - \rm{{F}_{20^{\circ}}} ) + \frac{\rm{{LE}_{CD}}}{\red{180^{\circ}}} + ( 1 - \rm{{LR}_{CD}} )}{4}.
    \label{eq:seld_error}
\end{align}

\subsection{Hyper-parameters}
\label{ssec:hyper}

The sampling frequency was set to 24 kHz.
The short-term Fourier transform~(STFT) was applied with 20 ms frame length and 10 ms frame hop.
Input features are segmented to have a fixed length of 1.27 seconds.
The shift length was set to 0.2 seconds during inference.
We used a batch size of~32, and each training sample was generated on the fly~\cite{erdogan2018investigations}.
We used the Adam optimizer with a weight decay of~$10^{-6}$.
We gradually increased the learning rate to 0.001 with 50,000 iterations~\cite{goyal2017accurate}.
After the warm-up, the learning rate was decreased by 10\% if the SELD error of the validation did not improve in 40,000 consecutive iterations.
We validated and saved model weights in every 10,000 iterations up to 400,000 iterations.
Finally, we applied stochastic weight averaging~(SWA)~\cite{izmailov2018averaging} to the last 10 models.
The threshold for activity is 0.5 to binarize predictions during inference.

\subsection{Experimental settings}
\label{ssec:setting}

We compared the multi-ACCDOA format with the single one.
The number of tracks in the multi-ACCDOA format was fixed at the maximum number of overlaps in the dataset, i.e., $N = 3$.
The difference in network architecture between the multi- and single-ACCDOA formats was only the final fully-connected layer.
We compared three PITs for the multi-ACCDOA format: the non-class-wise PIT, the class-wise PIT, and the class-wise ADPIT.
We also investigated the threshold of unification from 15 to 45 degrees for inference.
Other configurations mostly followed our DCASE 2021 challenge settings~\cite{shimada2021ensemble} in all methods.
Multichannel amplitude spectrograms and inter-channel phase differences~(IPDs) were used as features.
Three data augmentation methods were applied: equalized mixture data augmentation~(EMDA)~\cite{takahashi2017aenet,takahashi2016deep}, rotation in the first-order Ambisonic~(FOA)~\cite{mazzon2019first}, and SpecAugment~\cite{park2019specaugment}.
A conventional CRNN~\cite{cao2019polyphonic} was used for the experiments.

The multi-ACCDOA format was compared with the track-wise method called EINV2~\cite{cao2021improved}.
We prepared the original EINV2 architecture~\cite{cao2021improved}.
Since the architecture uses MHSA blocks, we also replace GRU layers with MHSA blocks for the multi-ACCDOA format.
We called the architecture CNN-MHSA.

Then we compared the proposed method with state-of-the-art SELD methods without model ensembles.
Although model ensemble techniques effectively achieve high performance, we omitted this for pure comparison between single models.
We used RD3Net~\cite{shimada2021accdoa} for comparison.
We used cosIPDs, and sinIPDs~\cite{wang2018multi}, instead of the IPDs for this model.
The STFT was applied with 40 ms frame length and 20 ms frame hop, and the input length was set to 5.11 seconds.

\subsection{Experimental results}
\label{ssec:res}

Table~\ref{tb:result_accdoa} summarizes the performance in the single- and multi-ACCDOA formats.
The multi-ACCDOA format with the class-wise ADPIT outperformed the single one for all metrics with only a few increases in parameters.
The multi-ACCDOA format with the class-wise ADPIT showed 7.2 points higher ${LR}_{CD}$ than the single one in the cases with overlaps from the same class, while the multi-ACCDOA format with the class-wise ADPIT performed comparably in the other cases.
The result shows the threshold value slightly affected the performance, especially ${LR}_{CD}$.
The result also shows that conventional PITs made it difficult to train the multi-ACCDOA format since the PITs cannot allow each track of the multi-ACCDOA format to learn with the same target as the single one.
It is also shown that the multi-ACCDOA format with the class-wise ADPIT can successfully localize and detect the overlaps from the same class and maintain the performance in the other cases.

Table~\ref{tb:result_einv2} summarizes the performance in the multi-ACCDOA format trained with the class-wise ADPIT and the track-wise format.
The multi-ACCDOA format performed better in $\rm{{ER}_{20^{\circ}}}$, $\rm{{F}_{20^{\circ}}}$, and $\rm{\mathcal{E}_{SELD}}$.
The result shows that the multi-ACCDOA format performed comparably to the track-wise format with considerably fewer parameters.

Table~\ref{tb:result_sota} shows the performances of the SELD methods without model ensembles.
The RD3Net using the multi-ACCDOA format trained with the class-wise ADPIT performed the best in $\rm{{ER}_{20^{\circ}}}$, $\rm{{F}_{20^{\circ}}}$, and $\rm{\mathcal{E}_{SELD}}$ for the DCASE 2021 Task 3 development set.

\begin{table}[t]
    \centering
    \caption{SELD performance of state-of-the-art methods and the proposed method evaluated for the testing fold of the development set. The threshold of unification is 30 degrees.}
    \scalebox{0.86}{
        \begin{tabular}{l|c|ccccc}
        \toprule
        & \# of & & & & & \\
        Method & params & $\rm{{ER}_{20^{\circ}}}$ & $\rm{{F}_{20^{\circ}}}$ & $\rm{{LE}_{CD}}$ & $\rm{{LR}_{CD}}$ & $\rm{\mathcal{E}_{SELD}}$ \\
        \midrule
        Shimada's~\cite{shimada2021ensemble} & 7.09 M & 0.436 & 67.9 & {\bf 11.5}$^{\circ}$ & 69.5 & 0.281 \\
        Nguyen's~\cite{nguyen2021dcase}      & 14.2 M & 0.450 & 70.0 & $13.1^{\circ}$ & {\bf 76.1} & 0.266 \\
        Baseline~\cite{politis2021dataset}   &  0.5 M & 0.73  & 30.7 & $24.5^{\circ}$ & 40.5 & 0.539 \\
        \midrule
        Ours                                 & 1.69 M & {\bf 0.413} & {\bf 70.3} & $13.3^{\circ}$ & 72.7 & {\bf 0.264} \\
        \bottomrule
        \end{tabular}
    }
    \label{tb:result_sota}
    \vspace{-4mm}
\end{table}

\section{Conclusion}
\label{sec:conclusion}

We extended the activity-coupled Cartesian direction-of-arrival~(ACCDOA) format to a multi one and proposed auxiliary duplicating permutation invariant training~(ADPIT).
The multi-ACCDOA format (a class- and track-wise output format) enables the model to solve the cases with overlaps from the same class.
The class-wise ADPIT makes each track of the multi-ACCDOA format be trained with original targets and duplicated targets.
The scheme enables each track to learn with the same target as the single one.
During inference, duplicated outputs are unified with their similarity.
In the evaluations on the sound event localization and detection task for DCASE 2021 Task 3, the model using the multi-ACCDOA format trained with the class-wise ADPIT detects overlapping events from the same class while maintaining the performance in the other situations.
The proposed method also performed comparably to state-of-the-art methods with fewer parameters.

\bibliographystyle{IEEEtran}
\bibliography{refs}

\end{document}